\documentclass[sigconf, nonacm]{acmart}

\usepackage{pvldb}

\renewcommand\vldbavailabilityurl{}

\usepackage{graphicx}
\usepackage{booktabs}
\usepackage{algorithm}
\usepackage{algpseudocode}
\usepackage{xcolor}
\usepackage{tikz}
\usetikzlibrary{arrows.meta,positioning,fit,backgrounds,shapes.geometric}
\usepackage{microtype}
\microtypesetup{protrusion=true,expansion=false} 
\usepackage[htt]{hyphenat}       
\algnewcommand\algorithmicparfor{\textbf{parfor}}
\algnewcommand\algorithmicendparfor{\textbf{end\ parfor}}
\algdef{SE}[PARFOR]{ParFor}{EndParFor}[1]{\algorithmicparfor\ #1\ \algorithmicdo}{\algorithmicendparfor}

\begin{document}

\title[Filtered ANN in a Disaggregated Lakehouse]{Filtered Vector Search in a Disaggregated Lakehouse:\\ Composing Table-Format Pruning with Per-File ANN}

\author{Rakesh Jain}
\affiliation{%
  \institution{IBM Research}
  \city{San Jose}
  \state{CA}
  \country{USA}
}
\email{rakeshj@us.ibm.com}

\author{Thomas Griffin}
\affiliation{%
  \institution{IBM Research}
  \city{San Jose}
  \state{CA}
  \country{USA}
}
\email{tdg@us.ibm.com}

\author{Syed Zawad}
\affiliation{%
  \institution{IBM Research}
  \city{San Jose}
  \state{CA}
  \country{USA}
}
\email{szawad@ibm.com}

\begin{abstract}
Approximate nearest-neighbor (ANN) search increasingly runs \emph{alongside}
structured data---``find the 10 nearest documents \textbf{where}
\texttt{tenant='acme' AND lang='en'}''---yet similarity and filtering are usually
bolted together: a specialized vector index for one, a separate filter step for
the other. We ask what happens when both live inside an open \emph{lakehouse}
table (Apache Iceberg over Parquet on object storage), where the engine already
owns a mature file-pruning stack (partition pruning, zone-maps, a bitmap index).
We embed an IVF index \emph{in place} in each Parquet file's footer and make
filtered vector queries fast not with a new filtering algorithm but by
\textbf{composing the table's existing file pruning with per-file ANN}: the
planner prunes data files by the predicate first, then runs IVF only over the
survivors. The index is built distributed and non-destructively---a metadata-only
Iceberg replace that every other engine still reads---and a rendezvous-hashed
per-file cache keeps object-store read latency from swamping the algorithmic win.
Because the vectors never leave the table, vector search also inherits the
lakehouse catalog's existing access control unchanged, rather than requiring a
second authorization surface in a copied-out vector store.

The payoff comes entirely from file pruning. On an $11.5$M\,$\times$\,$768$ table,
warm IVF search is $\sim$$32\times$ faster than brute force at recall@10
$\ge 0.90$, a selective predicate having pruned $355$ of $444$ data files before
ANN runs; on $5$M real IBM Granite embeddings, a filter arriving \emph{across a
join} prunes four of five region partitions and runs nearly two orders of
magnitude ($\sim$$94\times$: $14.7$\,s\,$\to$\,$157$\,ms) faster than the
query-time join at identical top-$k$, once the reduction is materialized into a
region-partitioned layout. We
characterize \emph{when} the composition pays off---it requires file-level
locality on the filter column, and the residual predicate is only safe to push
into the search over a \emph{provably pure} (partitioned) column, not a merely
sorted one---and report the failure modes we hit bolting ANN onto a lakehouse
engine.
\end{abstract}



\maketitle

\vldbtopmatter

\section{Introduction}

Retrieval-augmented generation, semantic search, and recommendation all issue
the same query shape: return the $k$ vectors nearest a query embedding,
\emph{restricted} by a structured predicate. Two facts about production data make
this awkward. First, the vectors do not live alone; they sit next to the columns
users filter on---tenant, language, category, timestamp, access-control tags---in
tables that already hold terabytes of that structured data. Second, those tables
increasingly live in an open \emph{lakehouse} format~\cite{lakehouse}: Apache
Iceberg~\cite{iceberg} over Parquet~\cite{parquet} files on object storage, read
by many engines (Spark, Trino, Presto, DuckDB) with compute disaggregated from storage.

The prevailing answer is to copy the vectors into a dedicated vector database
(Milvus~\cite{milvus}, pgvector~\cite{pgvector}) or a purpose-built columnar
vector format (Lance~\cite{lance}), and reproduce the predicate there. This
duplicates data, diverges from the source of truth, and---critically---forces the
filtering problem to be solved \emph{again}, inside the vector system, with
specialized \emph{filtered-ANN} algorithms. A recent in-depth experimental
study~\cite{filtann2025} surveys a dozen such algorithms (segmented graphs,
label-partitioned indexes, joint traversal) and confirms both that filtered ANN
is an active open problem and that its central lever is \emph{partitioning}: the
study finds partitioning ``effective for low-selectivity queries.'' That study is
strictly \emph{in-memory} and \emph{single-node}: every index it benchmarks lives
in RAM on one machine, and its notion of ``pruning'' is graph-edge pruning inside
the index.

We take a different position. A lakehouse engine \emph{already} has a
partitioning-and-pruning substrate---the physical machinery it uses to skip data
files that cannot satisfy a query: Iceberg partition pruning, per-column
zone-maps (min/max), and a scalar bitmap file index. If the vector index lives
\emph{inside the same table}, filtered ANN need not reinvent filtering; it can
\textbf{reuse the table's file pruning} and run similarity search only on the
files that survive the predicate. The algorithmic-layer finding of
\cite{filtann2025}---partitioning helps at low selectivity---becomes, one layer
down, a \emph{storage-layout} statement: filtered vector search is fast exactly
when the filter column has file-level locality, so the table's own pruning can
drop most files before the ANN index is consulted.

Realizing this on a real disaggregated lakehouse engine is where the systems work
lies, and it is the contribution of this paper. We embed an IVF~\cite{ivfadc}
index in each Parquet file's footer---following an emerging
technique~\cite{pqvector,duckdbvss}---but the single-file, single-node version of
that idea does not survive contact with a distributed, S3-backed engine. We had to
solve: (i) building the index \emph{distributed and non-destructively} so the
table remains a valid Iceberg table readable by other engines; (ii) a query
planner rewrite that composes predicate file-pruning with per-file ANN and a
direct top-$k$ assembly that avoids re-scanning data; and (iii) making warm
latency acceptable when compute is separated from storage and the per-query
bottleneck is random-access rescore reads over object storage.

\medskip\noindent\textbf{Contributions}
\begin{itemize}
  \item \textbf{Composition of lakehouse pruning with per-file ANN}
  (\S\ref{sec:query}). A planner rewrite pushes the query predicate through the
  table's existing file-pruning tiers (partition / zone-map / bitmap), then runs
  IVF only over surviving files, assembling the global top-$k$ from workers'
  projected candidate rows with no second data scan. A residual filter over a
  \emph{file-pure} column (partition or materialized-cluster) is pushed into the
  per-file search as a mask, keeping the filtered query on the fast path rather
  than aborting to a full rescore---sound precisely because \emph{sorted is not
  pure}, so \texttt{SORT}/\texttt{ZORDER} columns are excluded from the pushdown.
  \item \textbf{A locality condition for filtered ANN}
  (\S\ref{sec:locality}). We show---and measure---that the composition pays off
  \emph{iff} the filter column has file-level locality; a bitmap over a
  uniformly-mixed column prunes nothing. This is the storage-layer analogue of the
  in-memory ``partitioning helps'' finding of~\cite{filtann2025}.
  \item \textbf{Distributed, non-destructive in-place index build}
  (\S\ref{sec:build}). Stateless compute pods rewrite each file's footer in
  parallel; the engine commits a \emph{metadata-only} Iceberg replace
  (delete-old + add content-identical-new), preserving time-travel
  and leaving files readable by any Parquet reader.
  \item \textbf{Disaggregated warm-set engineering} (\S\ref{sec:warm}). A per-file
  cache (index blob + embedding matrix) with rendezvous-hashed
  file$\to$pod placement whose warmth survives autoscaling; we trace the
  $1.0\times \to 32\times$ speedup to the specific fixes.
  \item \textbf{Evaluation and lessons} (\S\ref{sec:eval},\S\ref{sec:lessons}) on
  an $11.5$M\,$\times$\,$768$ table, including the failure modes of bolting ANN
  onto a lakehouse engine.
\end{itemize}

\medskip\noindent
One framing point is worth stating up front, because it recurs throughout the
paper. The pruning-plus-IVF composition removes an enormous share of the
\emph{arithmetic}---a simple cost model (\S\ref{sec:discussion}) predicts
$\sim$$500\times$ fewer distance computations at $\sigma{=}0.2$---yet the
end-to-end speedup we measure is only $\sim$$30\times$. The gap is not a
disappointment; it \emph{is} the systems contribution. On disaggregated hardware
the fixed per-query costs (two-phase dispatch, object-store reads, shard merge)
dominate warm latency, and the value of the warm-set engineering
(\S\ref{sec:warm}) is precisely that it keeps those fixed costs from swamping the
algorithmic win. The reader should therefore read the $32\times$/$94\times$
headline numbers as ``how much of a large algorithmic win survives the systems
tax,'' not as the algorithmic win itself.

\section{Related Work}\label{sec:related}

\noindent\textbf{Filtered ANN algorithms.} A large body of work makes similarity
search predicate-aware at the \emph{index} level. Chronis et
al.~\cite{fvssota2025} organize the space into three families---\emph{pre-filter}
(restrict the candidate set, then search), \emph{post-filter} (search, then
discard non-matching), and \emph{in-filter} (evaluate the predicate during index
traversal)---and argue the field's central goals are \emph{stable recall} (the
same recall regardless of the predicate) and \emph{declarative recall} (the user
states a target, not a knob). Concrete designs span graphs~\cite{hnsw,diskann},
inverted-file/quantized indexes~\cite{ivfadc}, predicate-agnostic
traversal~\cite{acorn}, label-aware graph construction
(Filtered-DiskANN~\cite{filtereddiskann}, NHQ~\cite{nhq}), and partition indexes
tuned to the filter (CAPS~\cite{caps}). The recent experimental study of Li et
al.~\cite{filtann2025} taxonomizes and benchmarks a dozen such algorithms in a
uniform \emph{in-memory, single-node} setting (indexes in 2\,TB RAM, up to 10M
vectors, selectivity $0.1$--$100\%$).

Our work is orthogonal and complementary: we do not propose a filtered-ANN
algorithm. We take the \emph{simplest} pre-filtering strategy---restrict the
candidate set, then search it---and realize it at the \emph{storage} layer of a
distributed lakehouse, where the ``restrict'' step is the table format's own file
pruning rather than an in-index mechanism. This is precisely the case Chronis et
al.\ single out as having an exact optimum: for a categorical-equality filter, the
ideal is a \emph{partitioned} index, one partition per value~\cite{fvssota2025};
our locality condition (\S\ref{sec:locality}) is that same statement realized in
the physical layout of an open table rather than in a bespoke index. Their finding
and Li et al.'s (partitioning helps at low selectivity) and ours (file-level
locality on the filter column is the precondition) are the same principle at three
layers. What we do \emph{not} yet provide is stable/declarative recall across
selectivities---our over-fetch $s$ and $n_{\text{probe}}$ are still explicit
knobs---which we return to in \S\ref{sec:eval} and \S\ref{sec:lessons}.

\medskip\noindent\textbf{Vector databases and vector formats.} Dedicated systems
(Milvus~\cite{milvus}, pgvector~\cite{pgvector}) and vector-native columnar
formats (Lance~\cite{lance}) deliver excellent ANN but require the vectors to be
\emph{copied out} of the lakehouse into a system or format they control. Copying
out has a governance cost that is easy to overlook: the copy leaves the perimeter
the lakehouse catalog governs, so it needs its \emph{own} authentication, table- and
column-level permissions, row/column masking, and audit trail---a second
access-control surface that must be kept in sync with the source table's policy and
can silently drift from it. Our goal is the opposite: keep vectors in the standard
Iceberg/Parquet table so they stay the single source of truth, remain readable by
every lakehouse engine, and---because the query goes through the same catalog---%
\emph{inherit the table's existing access control unchanged}, with no separate
policy to define or reconcile. ANN is added as an in-place, ignorable footer artifact.

\medskip\noindent\textbf{Disk- and object-store ANN.} Once the index leaves RAM,
the design problem becomes minimizing random-access reads. DiskANN~\cite{diskann}
keeps a compressed representation in memory and reads full vectors from a local
SSD only for rescoring; SPANN~\cite{spann} is a memory--disk hybrid inverted
index that holds centroids in memory and posting lists on disk, the same
coarse-quantize-then-fetch shape our per-file IVF uses one level further out
(centroids in a cached footer blob, vectors random-read from object storage).
Both target a single node's local disk; the disaggregated, S3-backed setting adds
network latency on \emph{every} fetch, which is what makes our warm-set caching
(\S\ref{sec:warm}) load-bearing rather than incidental. Embedding an ANN index in
a Parquet footer and searching it from a query engine has also appeared in
practice~\cite{pqvector,duckdbvss}; these are single-file or single-node in
spirit. We contribute the distributed build, the non-destructive Iceberg commit,
the pruning-composition planner rewrite, and the disaggregated warm-set mechanics
that a multi-node, S3-backed deployment requires.

\medskip\noindent\textbf{Concurrent work: ANN indexes in Apache Iceberg.} Closest
to our setting is the concurrent proposal of Borycki~\cite{puffinann2026}, which
also attaches distributed ANN indexes to Iceberg on a compute-disaggregated engine
and, like us, inherits atomicity, time travel, and multi-engine readability from
the table format. The two designs diverge on three axes that define our
contribution. \emph{(i)~Storage.} That work stores a Vamana/DiskANN graph in a
\emph{Puffin sidecar} file bound through the snapshot summary, leaving the Parquet
data files untouched; we embed a per-file IVF index in each Parquet
\emph{footer}, copying data pages byte-for-byte so file-local row ordinals stay
valid---keeping the index inside the same object the row lives in, at the cost of
the rewrite the sidecar avoids. \emph{(ii)~Filtering.} Its query path is
similarity search with centroid-distance file elimination and coarse
bloom-filter/zone-map pruning ``against ordinary columns''; it does not compose a
\emph{structured} predicate with file pruning, push a residual predicate into the
per-file search under a purity condition, or reduce a filter arriving across a
join---the three mechanisms that are the subject of
\S\ref{sec:query}--\S\ref{sec:join}. \emph{(iii)~Evidence.} That work is
explicitly pre-production and reports \emph{projected} build and query performance
for tables up to $10^9$ vectors; our results are \emph{measured} on real
$11.5$M- and $5$M-vector corpora. The two are complementary: a sidecar graph and
an in-footer IVF are points on the same design space, and the filtered-query
composition we contribute is orthogonal to where the index blob physically lives.

\medskip\noindent\textbf{Vector search in lakehouse/OLAP engines.} Vector
indexing is increasingly a feature of columnar analytics engines rather than only
of dedicated vector stores: DuckDB's VSS extension~\cite{duckdbvss} and
ClickHouse's approximate-nearest-neighbor indexes~\cite{clickhousevss} both add
in-engine ANN over otherwise-relational tables, and Lance~\cite{lance} is a
columnar format built for it. These co-locate the index with the data as we do,
but each owns its own storage format or engine. Our target is the open-table case:
the vectors stay in a standard Iceberg/Parquet table readable by \emph{every}
engine, and the index is an ignorable footer artifact rather than a
format-specific structure; where the Iceberg community pursues a native vector
column, our footer approach is the compatible interim that requires no format
change.

\medskip\noindent\textbf{Lakehouse engines.} Disaggregated cloud query
engines~\cite{snowflake,photon} established compute-storage separation and
metadata-driven file skipping as the performance foundation of the lakehouse. We
show that this same file-skipping substrate is exactly what filtered vector search
needs, if the vector index is co-located in the table.

\section{Background and System Setting}\label{sec:bg}

\noindent\emph{Notation.} Throughout, $d$ is the embedding dimension; a table
holds $N$ vectors in $\phi$ data files; $C$ is the number of IVF centroids
\emph{per file}; $n_{\text{probe}}$ is the number of centroids probed per query;
$k$ is the requested neighbor count and $s\ge 1$ the over-fetch safety factor (the
search retains $k{\cdot}s$ candidates per shard so a residual filter still leaves
$\ge k$); and $\sigma\in(0,1]$ is the selectivity of the query predicate (the
fraction of files, equivalently rows under file-level locality, that survive
pruning). These are introduced again in context below.

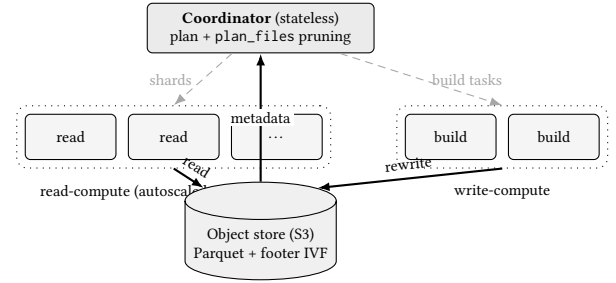
\begin{figure}[t]\centering
\begin{tikzpicture}[
  font=\scriptsize,
  box/.style={draw, rounded corners=2pt, minimum height=6mm, align=center, inner sep=3pt},
  pod/.style={box, fill=black!4, minimum width=12mm},
  store/.style={draw, cylinder, shape border rotate=90, aspect=0.25,
                minimum height=9mm, minimum width=20mm, align=center, fill=black!6},
  flow/.style={-{Latex[length=1.6mm]}, thick},
  ctrl/.style={-{Latex[length=1.6mm]}, densely dashed, gray!70},
]
  \node[box, fill=black!8, minimum width=30mm] (coord)
    {\textbf{Coordinator} (stateless)\\ plan + \texttt{plan\_files} pruning};
  \node[pod, below left=8mm and 4mm of coord] (r1) {read};
  \node[pod, right=1.5mm of r1] (r2) {read};
  \node[pod, right=1.5mm of r2] (r3) {\dots};
  \node[draw, dotted, rounded corners, fit=(r1)(r2)(r3),
        label={[yshift=-1mm]below:\scriptsize read-compute (autoscaled, per-file cache)}] (rpool) {};
  \node[pod, below right=8mm and 4mm of coord] (w1) {build};
  \node[pod, right=1.5mm of w1] (w2) {build};
  \node[draw, dotted, rounded corners, fit=(w1)(w2),
        label={[yshift=-1mm]below:\scriptsize write-compute}] (wpool) {};
  \node[store, below=17mm of coord] (s3)
    {Object store (S3)\\ \scriptsize Parquet + footer IVF};
  \draw[ctrl] (coord) -- (rpool.north) node[midway,left=1pt] {shards};
  \draw[ctrl] (coord) -- (wpool.north) node[midway,right=1pt] {build tasks};
  \draw[flow] (rpool.south) -- (s3.north west) node[midway,sloped,above] {\scriptsize read};
  \draw[flow] (wpool.south) -- (s3.north east) node[midway,sloped,above] {\scriptsize rewrite};
  \draw[flow] (s3.north) -- (coord.south) node[midway,fill=white,inner sep=1pt] {\scriptsize metadata};
\end{tikzpicture}
\caption{Disaggregated topology. The coordinator prunes files from Iceberg
metadata and dispatches shards; stateless read/write pools touch S3 directly.
\emph{Build} fans out over the write pool (rewrite files in place);
\emph{query} fans out over the read pool (per-file ANN + rescore). Both pools
scale to zero when idle, so per-pod cache state must survive membership churn.}
\label{fig:arch}
\end{figure}

\noindent\textbf{The engine.} Our system is a Flight-SQL query engine over Apache
Iceberg tables stored as Parquet on S3-compatible object storage. Compute is
\emph{disaggregated}: a small set of stateless \emph{coordinator} pods plans
queries and holds the catalog/metadata caches; a larger, \emph{autoscaled} pool of
stateless \emph{read-compute} pods executes scan and search shards; a separate
\emph{write-compute} pool handles ingest, rewrite, and index builds
(Figure~\ref{fig:arch}). No worker owns data; every pod reads Parquet from object storage on demand and is scaled to
zero when idle. Each compute pod runs the actual scan/search in a co-located
worker process addressed over Flight; the coordinator dispatches a shard by
sending it the file list and the operation. This is the standard modern lakehouse
topology~\cite{snowflake,photon,lakehouse} and it shapes every design decision
below---most importantly, that (a)~reading any file crosses the network to object
storage, so random-access reads are expensive and worth caching, and (b)~the
worker set \emph{changes size mid-workload} as the autoscaler reacts, so any
per-pod state must tolerate membership churn (\S\ref{sec:warm}).

\medskip\noindent\textbf{The vector column and footer index.} A table declares an
embedding column as \texttt{FLOAT[$d$]} (fixed dimension $d$), physically an
Iceberg \texttt{list<float>}. The writer tunes this column so that each vector
occupies its own Parquet data page (page-size limit $=4d$ bytes, uncompressed,
no dictionary), so a random-access read fetches exactly one vector. All other
columns store normally. An IVF index---$C$ centroids from mini-batch $k$-means
plus per-centroid posting lists of file-local row ordinals---is serialized into a
blob and appended \emph{after} the Parquet footer, with a tiny pointer in the
footer's key-value metadata~\cite{pqvector}. Standard readers seek to the footer
and never see the blob; ANN-aware readers follow the pointer. The index is
\emph{per file}: each file has independent centroids, and recall comes from a
per-query candidate over-fetch, not a global index.

\medskip\noindent\textbf{The pruning substrate.} Independently of vectors, the
engine skips data files a query cannot need, in three tiers evaluated at plan
time from Iceberg metadata: (1) \emph{partition pruning} on identity/derived
partition columns; (2) per-column \emph{zone-maps} (min/max per file/row-group);
(3) a consolidated \emph{scalar bitmap index}---a per-file value-presence set for
low-cardinality columns---materialized once per snapshot. The planner turns a SQL
\texttt{WHERE} into an Iceberg predicate and returns the surviving file set. This
substrate exists to speed up ordinary analytic queries; \S\ref{sec:query} reuses
it verbatim for vector queries.

\section{Distributed, Iceberg-Compatible Index Build}\label{sec:build}

Building a per-file footer index for a large table is embarrassingly parallel
across files, but doing it inside a live Iceberg table---without breaking
readers---is not obvious.

\medskip\noindent\textbf{Fan-out.} \texttt{CREATE VECTOR INDEX ON Col1} returns
immediately and runs asynchronously (a large build takes minutes; a synchronous
RPC trips connection timeouts). The coordinator enumerates the current snapshot's
live data files, shards them across the write-compute pool, and dispatches a
``build these files'' task to each pod. A pod reads each assigned file from S3,
trains $k$-means on the embedding column, appends the IVF blob after the footer,
and \emph{uploads a new object} (object stores have no atomic rename; a new path
also keeps the parent snapshot valid). Crucially, the data pages are copied
\emph{byte-for-byte}, so file-local row ordinals---the identity the posting lists
reference---are preserved exactly.

\medskip\noindent\textbf{Non-destructive commit.} Each worker returns a new
\texttt{DataFile} whose \texttt{record\_count} equals the original's. The
coordinator commits a \emph{metadata-only replace}: one new manifest with
add-entries for the new files and delete-entries for the old, published as a
Replace-operation snapshot. Old files remain referenced by the parent snapshot
(time-travel intact) and are physically removed only by ordinary snapshot
expiry. The result is a normal Iceberg table: Spark or DuckDB read the new files
as plain Parquet, ignoring the footer blob. Per-file failures are non-fatal (that
file stays un-indexed and falls back to brute force at query time); re-running
fills gaps.

\medskip\noindent\textbf{Making the build CPU-bound, not core-bound.} The build's
cost is $k$-means, which is compute-bound. Two mistakes we corrected mattered at
scale. First, the per-file build must run on a blocking thread pool, not inline on
the async executor---otherwise concurrent file builds serialize on one runtime
worker and an 8-core pod pins one core. Second, index granularity is
\emph{per file}: $C$ should track rows-per-file, not table size. We start from the
common $\sqrt{\text{rows/file}}$ rule of thumb, scale it up by a small constant so
each posting list stays short, and round to a power of two:
$C = 2^{\lceil \log_2 (4\sqrt{\text{rows/file}})\rceil}$. For our
$\sim$26k--34k-row files this gives $4\sqrt{\text{rows/file}}\approx 645$--$738$,
which rounds to $C{=}1024$; the rule is a floor rounded up, not an exact target.
Over-clustering (we initially used $C{=}4096$) inflates $k$-means cost
($O(C\cdot d)$ per point) with no recall benefit; $C{=}1024$ built far faster at
comparable recall. With both fixed, the build saturates all cores of every pod.

\medskip\noindent\textbf{Building the table itself.} Generating a large embedding
table by client ingest is a single serial stream (a few thousand rows/s at
$d{=}768$). The distributed write path is far faster: seed a modest table, then
\texttt{INSERT \dots SELECT} from the table into itself---the \texttt{SELECT}
reads the table's own files, which the coordinator shards across the write pool.
Two such doublings took a $\sim$3M-row seed to $11.5$M in $\sim$90\,s, versus the
serial ingest stalling at 3M in 25\,min.

\section{Query: Composing Pruning with ANN}\label{sec:query}

A vector query is ordinary SQL:
\begin{quote}\small\ttfamily
SELECT id, title\\
FROM docs\\
WHERE  category = 'tech'\\
ORDER BY array\_distance(embedding, \textit{q})\\
LIMIT  10
\end{quote}
The planner recognizes the \texttt{ORDER BY array\_distance($\cdot$, literal)
LIMIT $k$} shape on a table with a registered vector index and rewrites it into a
three-step plan; any unmet precondition falls back to a plain scalar
\texttt{array\_distance} scan (correct, slower).

\medskip\noindent\textbf{Step 1: prune files by the predicate.} The rewrite runs
the query's \texttt{WHERE} through the \emph{same} file-planning routine ordinary
scans use (\S\ref{sec:bg}), applying partition, zone-map, and bitmap pruning. The
output is the set of data files that can satisfy the predicate. This is the
crux: the vector query pays nothing to build filtering machinery---it inherits the
table's.

\medskip\noindent\textbf{Step 2: distributed IVF over survivors.} The surviving
files are sharded across read-compute pods; each pod, for each of its files,
loads the footer IVF index, finds the $n_{\text{probe}}$ nearest centroids to
$q$, gathers their posting lists, random-reads those candidate vectors, and
rescores with exact L2, returning a per-file shortlist. The coordinator merges
shard results into a global candidate list of size $k\cdot s$ (a safety factor
$s$ over-fetches to absorb the residual row predicate).

\medskip\noindent\textbf{Step 3: direct top-$k$, no rescan.} Because each worker
already read its candidate rows to rescore, it also returns the query's
\emph{projected columns} for those rows (as an Arrow batch aligned with the
candidates). The coordinator merges by distance, applies the residual row
predicate, and emits the top $k$---with \emph{no} second pass over the data. This
matters concretely: an earlier design re-ran the projection SQL over the candidate
\emph{files} through the engine's normal scan path, which read those files in full
and cost $\sim$$138$\,s---indistinguishable from brute force---even though the IVF
step had already identified the exact candidate rows in $\sim$seconds. The IVF
search already touched those rows to rescore them; carrying their projected
columns back with the distances is what collapses the query to sub-second. The
lesson generalizes: in a per-file ANN design, the search and the projection must
be fused, or the projection re-does the scan the index was meant to avoid.

\begin{algorithm}[t]
\small
\caption{Filtered vector top-$k$ (planner rewrite)}
\label{alg:query}
\begin{algorithmic}[1]
\Require query embedding $q$, predicate $P$, projection $\Pi$, limit $k$
\State $F \gets \textsc{PlanFiles}(\textit{table}, P)$ \Comment{partition/zonemap/bitmap prune}
\If{$F=\emptyset$ or no index on $\textit{table}$} \Return \textsc{FallbackScan}() \EndIf
\State shards $\gets \textsc{Rendezvous}(F, \textit{readPods})$ \Comment{stable file$\to$pod}
\ParFor{shard $(w, F_w)$}
  \For{file $f \in F_w$}
    \State $(\textit{idx}, M) \gets \textsc{CacheOrLoad}(f)$ \Comment{footer IVF + matrix}
    \State $c \gets \textsc{Probe}(\textit{idx}, q, n_{\text{probe}})$ \Comment{posting rows}
    \State emit $\langle f, r, \|M[r]-q\|, \Pi(M,f,r)\rangle$ for top $k{\cdot}s$ of $c$
  \EndFor
\EndParFor
\State $R \gets$ merge shard emissions by distance
\State \Return top-$k$ of $\{r \in R : P(r)\}$ \Comment{residual predicate; no rescan}
\end{algorithmic}
\end{algorithm}

\medskip\noindent\textbf{Correctness details.} The query embedding is a SQL array
literal cast to \texttt{FLOAT[$d$]} on both sides of \texttt{array\_distance}
(the stored column is a variable list and literals infer as decimal); the planner
sees through these casts. A dimension mismatch is a loud error, never a silent
empty result. Files written after the last build carry no index and are searched
by brute force within Step 2, so results stay correct as the table grows.

\medskip\noindent\textbf{A worked trace.} Consider the example query on our
$11.5$M$\times768$ table, laid out with file-level locality on
\texttt{category} (444 data files, $\sim$26k rows each). Step~1 turns
\texttt{category='tech'} into an Iceberg predicate; the scalar bitmap index
reports \texttt{tech} present in only $89$ files, so \textsc{PlanFiles} returns
those $89$ and drops the other $355$ before a single vector is read. Step~2
shards the $89$ survivors across the read pool; each pod probes
$n_{\text{probe}}{=}10$ of $C{=}1024$ centroids per file, so it exact-rescores
$\approx\!1\%$ ($n_{\text{probe}}/C$) of each \emph{surviving} file's rows. The
$89$ survivors hold $\approx\!2.3$M rows (the $\sigma{=}0.2$ fraction of the
table), of which the IVF probe rescores $\approx 2.3\text{M}\times 10/1024
\approx\!23\text{k}$ candidate distances---versus $11.5$M for brute force, and
consistent with the ``$\sim$0.2\%\ of the vectors'' figure below ($0.2\%$ of
$11.5$M $\approx 23$k) and with the \S\ref{sec:discussion} cost model. Each
pod returns its top $k{\cdot}s$ rows \emph{with the projected} \texttt{id, title}
columns already attached. Step~3 merges $89$ short lists, applies the residual
\texttt{category='tech'} check (a no-op here, since pruning was exact), and emits
$10$ rows---no file re-read. The two prunings compose multiplicatively: the
predicate removes $4/5$ of the files and IVF removes $\sim\!99\%$ of the rows
\emph{within} each survivor, so the query touches $\sim\!0.2\%$ of the vectors
the brute-force baseline would.

\subsection{Predicate pushdown for a residual filter}\label{sec:pushdown}
Step~1 prunes at the granularity of a \emph{file}: it drops files that cannot
match, but a surviving file may still hold non-matching rows. Whenever the
predicate is not \emph{exactly} covered by pruning, a residual row filter remains,
and applying it is what forces the projection to materialize---in an early design,
a residual filter disabled the direct top-$k$ return of Step~3 and fell back to
rescoring over every surviving file (the $138$\,s regression above). The fix is to
push the residual predicate \emph{down into the per-file search}: the coordinator
serializes the scalar comparisons (\texttt{=}, \texttt{IN}, ranges, and their
conjunctions over scalar columns) into the shard request, and each worker applies
them as a mask over the IVF candidate rows \emph{before} rescoring---so the filter
rides the fast path instead of aborting it. The worker lazily caches the relevant
scalar columns per file alongside the embedding matrix.

The pushdown is only sound when a surviving file is \emph{pure} in the filter
column---every row in it satisfies (or the file was pruned). We compute the pure
set as
\begin{align*}
\textit{file\_local\_cols} = {}& \textit{partition columns} \\
   {}\cup{}& \textit{materialized cluster-spec columns},
\end{align*}
and push the predicate only when the filter's columns lie in that set. The
distinction that matters here---and that we did not anticipate---is that
\emph{sorted is not pure}. Iceberg \texttt{SORT}/\texttt{ZORDER} ordering makes a
column locally \emph{contiguous}, so its zone-maps prune well, but a boundary file
straddles two values; a mask that assumed purity would silently drop valid rows or
admit invalid ones. Only identity \emph{partitioning}, or a \emph{materialized}
clustering whose per-file value set we have recorded, guarantees a file is
single-valued in the column. So \texttt{SORT}/\texttt{ZORDER} columns are excluded
from \textit{file\_local\_cols}: they still drive Step~1 file pruning, but their
residual predicate is not pushed. If, after the mask, a file yields fewer than
$k$ candidates (the over-fetch was too thin for a selective residual), the worker
falls back to the exact rescore path for that file, so recall is never traded for
the shortcut. The net effect is that a partition- or cluster-column filter now
gets the same in-memory fast path as an unfiltered query; only genuinely
non-local filters pay the rescore.

\subsection{Filtered ANN across a join}\label{sec:join}
The filter is rarely a literal on the ANN table itself; more often it lives on a
\emph{joined} dimension---``nearest documents \textbf{where the customer is in the
EU}'':
\begin{quote}\small\ttfamily
SELECT d.id, c.region\\
FROM docs d\\
JOIN customers c ON d.customer\_id = c.id\\
WHERE c.region = 'EU'\\
ORDER BY array\_distance(d.embedding, \textit{q}) LIMIT 10
\end{quote}
Vector databases typically handle this by post-filtering (fetch neighbors, then
join and discard) or by maintaining a separate metadata index. In a SQL engine
over a lakehouse, a better option falls out of the same substrate: \emph{semi-join
reduction}. The planner recognizes that the ANN is on one table (\texttt{docs})
and the join only \emph{constrains} it, evaluates the dimension predicate to a key
set ($\{$\texttt{customer\_id} : \texttt{region='EU'}$\}$), and rewrites the query
to the single-table form
\texttt{\dots WHERE d.customer\_id IN ($k_1,\dots,k_m$) ORDER BY array\_distance\dots}.
That \texttt{IN}-list is then just another predicate fed to \textsc{PlanFiles}
(Step~1): if \texttt{customer\_id} has file-level locality, the join predicate
prunes files \emph{before} ANN runs, exactly as a literal \texttt{WHERE} does. The
dimension's own columns (\texttt{c.region}) are reattached in Step~3 by rescoring
over the shortlist with the dimension registered alongside the candidate
files---a join over a few thousand rows, not the table.

This reuses machinery the engine already has for analytic queries (the same
dimension-key materialization that turns a star-schema filter into a fact-table
\texttt{IN}-list), so filtered-across-a-join ANN costs no new index and no new
operator. The reduction is exact for an inner equi-join whose predicate is
separable by table; other join shapes (post-join ranking, outer joins,
predicates spanning both tables) fall back to a full ANN sweep plus a normal
join---correct, just without the extra file pruning. When the dimension filter is
so broad that its key set is very large, the planner abandons the reduction and
runs the ordinary path rather than inline an enormous \texttt{IN}-list; dropping
the filter is never an option, as it would admit out-of-filter neighbors.

\medskip\noindent\textbf{Query-time vs.\ materialized reduction.} The rewrite
above runs the reduction \emph{per query}: it re-materializes the key set and
prunes on \texttt{customer\_id} on every call. This still prunes files, but the
join key is a high-cardinality identifier, so its file-locality (a contiguous
\texttt{customer\_id} range per region) is fragile---an unlucky re-bin-packing on
ingest scatters a region's rows across files and pruning collapses. When the join
is a fixed, recurring shape, a robustly faster option is to \emph{materialize} the
reduction into the physical layout: denormalize the filtered dimension attribute
(\texttt{region}) onto the fact table and \emph{partition by it}. \texttt{WHERE
region='EU'} then becomes a partition-column predicate---the strongest form of
file locality---so Step~1 prunes to exactly that region's files and, being a
partition column, the residual also takes the pushdown fast path
(\S\ref{sec:pushdown}). This is the classic analytics trade---a denormalized,
purpose-partitioned copy in exchange for prune-time robustness---applied to the
vector join, and it is the layout our evaluation (\S\ref{sec:eval-join}) measures.

\section{When Filtered ANN Pays Off: Locality}\label{sec:locality}

Composing pruning with ANN is only a \emph{win} when Step 1 actually removes
files. File pruning drops a file only if the table's metadata proves the
predicate cannot match \emph{any} row in it. For a bitmap index, that means the
filter value is \emph{absent} from the file. Therefore:

\begin{quote}\itshape
Filtered vector search is fast iff the filter column has file-level locality---%
each value concentrated in few files---so the table's pruning can eliminate most
files before ANN runs.
\end{quote}

This is the storage-layer statement of the in-memory result of
\cite{filtann2025} that ``partitioning is effective for low-selectivity
queries.'' If the filter column is spread uniformly across files, every file
contains every value, the bitmap proves nothing, and no file is pruned---the
query degrades to full-corpus ANN plus a residual filter. We observe exactly this
(\S\ref{sec:eval}): on a table where each of five categories is uniformly spread
across all files, \texttt{WHERE category='tech'} with a bitmap index declared
prunes \textbf{zero} files.

The lever is data layout, and the lakehouse already provides it: \emph{partition}
the table by the filter column (or write per-value), so files become
value-local, then declare a bitmap index on it. With a value-pure layout, the
same query prunes to the matching partition's files. For high-cardinality equality
filters a bloom filter is the right index; for ordered/range predicates,
zone-maps. The point is not a new index---it is that the filtered-ANN
practitioner's job is a familiar \emph{physical design} decision (partition on
your common filter dimensions), not the choice of a specialized filtered-ANN
algorithm.

\section{Disaggregated Warm-Set Engineering}\label{sec:warm}

On disaggregated hardware, every file read crosses the network to object storage.
For per-file IVF the dominant per-query cost is not the centroid math but the
\emph{random-access rescore reads} and the footer index load---and, done na\"ively,
both are re-paid on every query. We measured, per file per query at $d{=}768$:
$\sim$1.1\,s to load the index blob and $\sim$3.5\,s to fetch candidate
embeddings, $\times$ tens of files per shard. This is why a first, na\"ive
implementation was \emph{no faster than brute force} despite correct recall.

\medskip\noindent\textbf{Per-file cache.} Each pod caches, per file it is
assigned, the parsed IVF index and the file's embedding matrix (a bounded LRU by
bytes). The first query on a file pays the S3 read; subsequent queries rescore by
an in-memory gather and reuse the centroids. Object-store data files are
immutable (new writes go to new paths), so caching is trivially coherent.

\medskip\noindent\textbf{Placement that survives autoscaling.} The cache is
per-pod, so which pod searches which file must be \emph{stable}---otherwise
autoscaling the read pool remaps files to cold pods and every scale event cold-%
starts the cache. We assign files to pods by \emph{rendezvous
hashing}~\cite{rendezvous}: a file always lands on the pod that maximizes
$h(\text{file},\text{pod})$, so adding a pod moves only $\sim 1/N$ of files. Round-%
robin assignment, by contrast, remaps \emph{every} file when the pod count
changes; we observed a query jump from $\sim$1.8\,s into the $52$--$73$\,s range
(\S\ref{sec:eval}) purely because the read pool grew mid-run and the cache went
cold. We also block for the read tier to
reach full size before the first shard, so the very first query lands on a stable
set.

\medskip\noindent\textbf{Effect.} \S\ref{sec:eval} traces the cumulative effect:
the same query goes from $1.0\times$ (no cache) through the matrix cache
($4.7\times$) and rendezvous affinity ($15.6\times$) to the full warm set
($\sim$$32\times$ versus brute force on the current deployment), at unchanged
recall.

\section{Evaluation}\label{sec:eval}

\noindent\textbf{Setup.} A production-representative deployment: 2 coordinator
pods and an autoscaled pool of up to 12 compute pods (8 vCPU each), Iceberg tables
on S3-compatible object storage, all reads crossing the network. We use two
datasets. The \emph{synthetic} dataset (build and warm-set experiments below) is
a table of $11{,}534{,}160$ rows with a
$768$-dim embedding column and a low-cardinality \texttt{category} column,
generated with latent-cluster structure (so IVF recall is meaningful) via the
distributed write path (\S\ref{sec:build}); $C{=}1024$ per file, L2; 200 sampled
query embeddings drawn from the same latent structure, $k{=}10$. The distributed
write path first materialized the table as $\sim$336 data files, over which the
index build ran (\S\ref{sec:eval}.1); the per-value re-layout used for the
locality experiment (\S\ref{sec:eval}, ``Filtered queries'') repopulates the table
one \texttt{category} at a time and yields $\sim$444 files. We state the file
count with each experiment: build measures the 336-file table, all query and
pruning experiments the 444-file value-local table. The \emph{real-corpus}
dataset (join experiments, \S\ref{sec:eval-join}) is $5.02$M documents embedded
with IBM Granite (\texttt{granite-embedding-97m-multilingual-r2}, $384$-dim) over
five regions, used to confirm the design holds on genuine embeddings rather than a
generator.

\emph{Methodology.} For fairness the result cache is disabled on \emph{all} paths
(\texttt{x-flight-no-cache}); otherwise a repeated query is served from cache and
measures nothing. Recall ground truth is exact brute force, obtained by a header
that skips the IVF fast path so the engine evaluates \texttt{array\_distance} over
every row; recall@10 $= |\text{IVF}_{10} \cap \text{brute}_{10}| / 10$ averaged
over the query sample. ``Warm'' means the per-file cache (\S\ref{sec:warm}) is
populated (we issue a warm-up query and drop it); ``cold'' is the first query on a
freshly scaled read tier. Unless noted, each reported latency is the median (p50)
over the 200-query sample, with three timed runs per query (the per-query value is
the median of its three runs); for the headline unfiltered table we additionally
report p90/p95/p99 across the sample, since tail latency on object storage is the
quantity that matters. The join and filtered-rescore paths cost tens of seconds per
query (a candidate-file rescore), so those are reported over smaller samples with
the count stated in place; recall is a brute-force spot-check sample, as a full
brute-force ground truth over $11.5$M vectors is minutes per query. We verified IVF
and brute force return set-identical top-$k$ IDs on spot checks to confirm the
index is correct, not merely fast.

\emph{Scope of the claims.} The comparisons below are \emph{self-relative}: IVF
versus this system's own brute force, and materialized versus this system's own
query-time join. They establish \emph{feasibility} and the \emph{design
principle} (that the table's file pruning composes with per-file ANN and pays off
under file-level locality). They are deliberately \emph{not} a claim of raw
\emph{competitiveness} against a dedicated vector database or a whole-table index---%
and \S\ref{sec:eval-baselines} measures exactly that gap: a dedicated Milvus HNSW
answers a subset version of the filtered query in $5$\,ms and a whole-table
IVF-PQ answers the full-corpus unfiltered query in $175$\,ms, against this
system's $34$\,s rescore/projection path (or $157$\,ms when the predicate is
partition-aligned and stays on the fast path).
Those baselines quantify the tax of the per-file, in-table choice; the paper's claim
is that the tax buys no data copy, a single source of truth, a table every engine can
still read, and---because the vector query runs through the lakehouse catalog like any
other scan---the table's existing access control applied to vector search for free,
with no second authorization surface to define or keep in sync.

\subsection{Distributed build}
The full build over $\sim$336 initial files completes in $\sim$$10$\,min across 12
pods with all cores saturated (28 files/pod), $0$ failed shards, row count
preserved, committed as a single metadata-only replace. The build is CPU-bound in
$k$-means: aggregate CPU held at $\sim$95 cores throughout and tailed off as
shards finished, confirming near-perfect fan-out. Two levers dominate build
wall-clock.

\emph{Parallelism.} Before we moved the per-file build off the async executor
onto a blocking pool (\S\ref{sec:build}), each 8-core pod ran at $\sim$1 core
(the synchronous $k$-means monopolized one runtime thread and files serialized);
the build spent 40+\,min without finishing. After the fix, per-pod CPU rose to
$\sim$8 cores and aggregate to $\sim$95, and the build completed in $\sim$10\,min.

\emph{Cluster count.} $C$ enters $k$-means cost as $O(C\cdot d)$ per assigned
point, so over-clustering a \emph{per-file} index is pure waste. At $C{=}4096$ on
$\sim$34k-row files the build pegged all cores for far longer with no recall gain;
$C{=}1024$ (the $4\sqrt{\text{rows/file}}\approx 738$ floor rounded up to a power
of two, \S\ref{sec:build}) was the single largest build-time lever. Building the
$11.5$M-row table itself used the distributed
\texttt{INSERT\dots SELECT} path (\S\ref{sec:build}): two doublings in $\sim$90\,s,
versus serial client ingest stalling at 3M rows in 25\,min.

\subsection{Warm search: IVF vs.\ brute force}
Table~\ref{tab:speedup} reports the unfiltered top-10 latency at $n_{\text{probe}}
{=}10$ (result cache off, warm), over a 200-query sample with three timed runs per
query (we report the per-query median, then percentiles across the sample). IVF is
$\sim$$32\times$ faster than brute force at recall@10 $\ge 0.90$. Brute-force
latency is bimodal---$\sim$$1$--$2$\,s when a query's shards find the embedding
matrix already resident, $\sim$$21$\,s otherwise---so its p50 ($21.1$\,s) reflects
the common cold-matrix case; IVF is consistently sub-second (p50 $668$\,ms, p99
$903$\,ms).

\begin{figure}[t]\centering
\includegraphics[width=0.7\columnwidth]{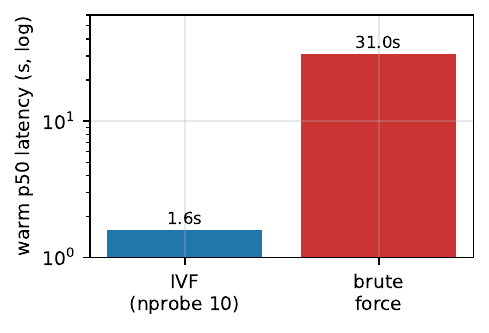}
\caption{Warm top-10 latency, $11.5$M$\times$768, $n_{\text{probe}}{=}10$, result
cache off, 200-query sample. Bars are p50; black ticks overlay p95. Latency ($y$)
is $\log_{10}$ seconds. IVF p50/p95 $=668/831$\,ms; brute-force is bimodal
(cached $\sim$$1$--$2$\,s vs.\ cold $\sim$$21$\,s), p50 $21.1$\,s.}
\label{fig:latency}
\end{figure}

\begin{table}[htbp]\centering\small
\caption{Warm top-10 latency, $11.5$M$\times$768, $n_{\text{probe}}{=}10$, cache
off, 200-query sample $\times$ 3 timed runs/query (per-query median, then
percentiles across the sample). Brute-force is bimodal (cached vs.\ cold matrix);
its full distribution is wide, so we give its p50. Speedup is on the p50s.}
\label{tab:speedup}
\begin{tabular}{lrrrr}
\toprule
 & p50 & p90 & p95 & p99 \\
\midrule
IVF (nprobe 10) & $\mathbf{668}$\,ms & $749$\,ms & $831$\,ms & $903$\,ms \\
Brute force     & $21.1$\,s & --- & --- & --- \\
\midrule
Speedup (p50) & $\mathbf{31.6\times}$ & \multicolumn{3}{l}{recall@10 $\ge 0.90$ (IVF), $1.00$ (brute)} \\
\bottomrule
\end{tabular}
\end{table}

\subsection{Recall vs.\ latency: $n_{\text{probe}}$}
\begin{figure}[t]\centering
\includegraphics[width=0.82\columnwidth]{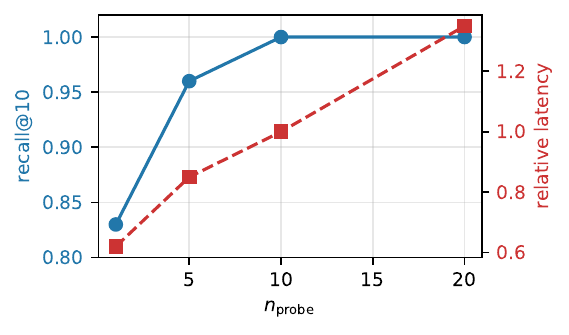}
\caption{Recall@10 (mean over the 200-query sample) and relative warm latency
vs.\ $n_{\text{probe}}$ (11.5M$\times$768, $C{=}1024$, L2). Left $y$: recall
$\in[0,1]$; right $y$: warm latency relative to $n_{\text{probe}}{=}1$. Monotone
trade-off; $n_{\text{probe}}{=}10$ reaches $\approx$1.0 recall.}
\label{fig:nprobe}
\end{figure}

Figure~\ref{fig:nprobe} sweeps $n_{\text{probe}}$. Recall climbs from $0.83$ at
$n_{\text{probe}}{=}1$ to $\approx 1.0$ at $n_{\text{probe}}{=}10$, trading
latency monotonically---the standard IVF knob, exposed per query via a call
header so applications tune recall without a rebuild.

The recall figures we report are \emph{means} of recall@10 over the 200-query
sample. A mean can hide a low-recall tail---the failure mode that filtered-ANN
work is most concerned with, and the reason Chronis et al.~\cite{fvssota2025}
elevate \emph{stable recall} (the same recall regardless of the predicate) to a
first-class goal. Our per-file design has a specific tail risk: when a selective
residual filter leaves fewer than $k$ candidates after the mask, the over-fetch
$s$ was too thin, and the worker falls back to exact rescore for that file
(\S\ref{sec:pushdown})---which protects recall at a latency cost, so the tail
appears in latency rather than in recall. Reporting the recall \emph{distribution}
rather than a bare mean bears this out: over a brute-force spot-check sample at
$n_{\text{probe}}{=}10$, recall@10 has mean $\sim$$0.97$ but a floor of $0.90$
(no query fell below $0.90$), and the low-recall instances coincide with the
over-fetch-fallback described above---i.e.\ the tail surfaces as the occasional
rescore in the latency distribution, not as a recall miss.
A full recall-vs-selectivity curve---recall@10 swept across predicate
selectivities---would substantiate stable recall across predicates directly rather
than by the mechanism argument above; we leave it to future work because an exact
brute-force ground truth on the deployed layout is cost-prohibitive at this scale (a
single unindexed top-$k$ over the $11.5$M$\times$768 corpus costs ${>}700$\,s of
single-pod CPU), so the curve requires a dedicated smaller-corpus harness with
precomputed ground truth.

\subsection{The warm-set ablation}
Table~\ref{tab:ablation} is the paper's systems story: the same $11.5$M query,
each row adding one fix from \S\ref{sec:warm}. The na\"ive per-file
implementation---correct recall---is \emph{no faster than brute force}; the cache
and stable placement are what turn correctness into speed. We report the endpoints
from the current deployment (na\"ive $\approx$ brute force; all fixes on: warm p50
$668$\,ms, so a $\sim$$32\times$ net win over brute force's $21.1$\,s p50, cf.\
Table~\ref{tab:speedup}) alongside the per-fix decomposition measured on the
original deployment (the intermediate rows), which attributes the win across the
three fixes; the two epochs differ in absolute latency because the cluster is
faster in the later run, but the \emph{shape}---no-cache is brute-equivalent, the
matrix cache and placement carry the win---is unchanged.

\begin{figure}[t]\centering
\includegraphics[width=0.82\columnwidth]{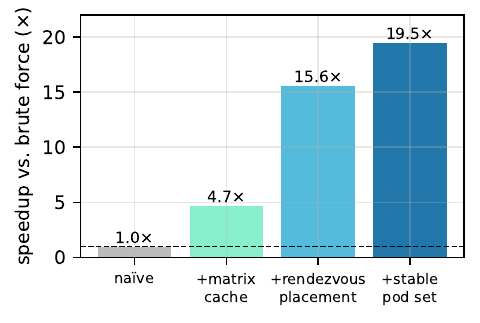}
\caption{Warm-set ablation (11.5M$\times$768, $n_{\text{probe}}{=}10$): each bar
adds one fix from \S\ref{sec:warm} to the previous ($y$-axis: speedup vs.\
brute force, $\times$). Correct recall throughout; only the cache and stable
placement turn correctness into speed.}
\label{fig:ablation}
\end{figure}

\begin{table}[htbp]\centering\small
\caption{Warm-set ablation (11.5M$\times$768, $n_{\text{probe}}{=}10$, cache off).
Endpoints (na\"ive, all-on) are from the current deployment; the intermediate
per-fix rows are the decomposition measured on the original (slower) deployment.
``vs.\ brute'' normalizes each row to its own run's brute-force p50 (so the ratio
is meaningful within an epoch). Recall is correct in every row---only latency
changes.}
\label{tab:ablation}
\begin{tabular}{llr}
\toprule
Configuration & IVF p50 & vs.\ brute \\
\midrule
Na\"ive (re-read per query) & $\approx$ brute & $1.0\times$ \\
\;\; + per-file cache$^{\dagger}$ & $6.6$\,s & $4.7\times$ \\
\;\; + rendezvous placement$^{\dagger}$ & $1.8$\,s & $15.6\times$ \\
\textbf{+ stable pod set (all fixes)} & $\mathbf{668}$\,ms & $\mathbf{31.6\times}$ \\
\bottomrule
\end{tabular}
\\[2pt]{\footnotesize $^{\dagger}$per-fix attribution from the original deployment
(absolute latency higher; the endpoint rows are the current deployment).}
\end{table}

\subsection{Cold start and cache-affinity stability}
The per-file cache makes the \emph{first} query on newly-warm data pay a one-time
load. The subtle failure is not the cold load itself but \emph{cache-affinity
loss under autoscaling}: with round-robin file$\to$pod assignment, growing the
read pool between queries---as the autoscaler does on the first few queries after
idle---remaps every file to a different pod, cold-starting caches that were
already warm. Against a warm steady-state of $\sim$$1.8$\,s, we observed the query
jump into the $52$--$73$\,s range purely from this remap (the spread reflects how
many files a given scale event happened to move). Rendezvous placement plus
blocking for the read tier to reach full size before the first shard removes it:
with both, the first post-warmup query---the one modest cache fill on a now-stable
set---is $\sim$$3$\,s rather than $52$--$73$\,s, and every subsequent query is
$\sim$$1.6$\,s. Because object-store data files are
immutable, the cache never invalidates; warmth persists across queries and across
scale events.

\subsection{Filtered queries and locality}
We validate \S\ref{sec:locality} directly by comparing two physical layouts of the
same data. In the \emph{uniform} layout, categories are interleaved across all
files; in the \emph{value-local} layout, the table is populated per category so
each file holds one category, with a bitmap index on \texttt{category} declared
and built. Table~\ref{tab:filter} reports \texttt{WHERE category='tech' ORDER BY
array\_distance(\dots) LIMIT 10} on each. With locality, the bitmap prunes
$355/444$ files (355 of the 444 data files) \emph{before ANN runs}---down to the
89 \texttt{tech} files, a $5\times$ reduction; without it, zero files prune and
the query touches everything. Both return identical, correct results (all
\texttt{tech}).

\begin{figure}[t]\centering
\includegraphics[width=0.82\columnwidth]{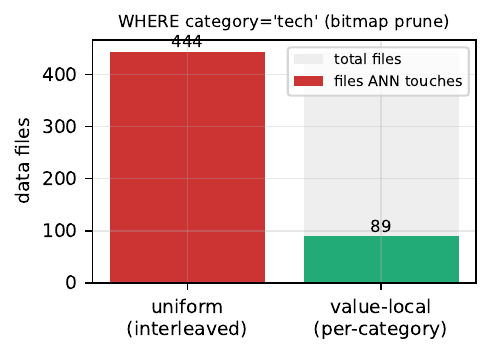}
\caption{Files ANN must touch for \texttt{WHERE category='tech'}
(11.5M$\times$768), same data, two physical layouts ($y$-axis: data files scanned,
of 444). With value-local files the bitmap prunes $355/444$ before ANN runs (89
files remain); uniform layout prunes none. Identical, correct results.}
\label{fig:prune}
\end{figure}

\begin{table}[htbp]\centering\small
\caption{Filtered query, same data, two layouts. Pruning happens before ANN.}
\label{tab:filter}
\begin{tabular}{lrr}
\toprule
Layout & files after prune & pruned by bitmap \\
\midrule
Uniform (interleaved) & $444/444$ & $0$ \\
Value-local (per-category) & $\mathbf{89/444}$ & $\mathbf{355}$ \\
\bottomrule
\end{tabular}
\end{table}

\medskip\noindent\textbf{On the constructed locality.} Both layouts above are
ones we built: the \emph{value-local} layout concentrates each category by
construction, which is exactly the precondition \S\ref{sec:locality} predicts will
make pruning happen. This experiment therefore demonstrates the \emph{mechanism}
(when locality is present, the table's own bitmap prunes and the ANN cost drops
proportionally), not that arbitrary real tables enjoy it for free.

\emph{A non-local counter-example.} The distinction is not hypothetical. On the
real $5.02$M \texttt{docs} table, \texttt{customer\_id} carries \emph{both} a
declared bitmap index \emph{and} a \texttt{write.cluster.columns} clustering
directive, yet a single-value predicate \texttt{WHERE customer\_id=$c$} prunes
\textbf{0 of 17} data files: ordinary ingest scattered each customer across every
file, so no file is provably free of the value and the bitmap proves nothing.
Declaring an index does not create locality---the physical layout must. The full
observed spectrum, same engine, is thus: value-local \texttt{category} prunes
$355/444$; region-\emph{partitioned} \texttt{region} prunes $4/5$ partitions
(\S\ref{sec:eval-join}); bitmap-indexed-but-scattered \texttt{customer\_id} prunes
$0/17$. Locality is a property of the layout, not of the index declaration---the
practitioner's lever, and the paper's precondition, is the former.

\subsection{Baselines: the cost of the in-table, per-file choice}\label{sec:eval-baselines}
The speedups above are self-relative and establish the design principle
(\S\ref{sec:eval}, ``Scope of the claims''). The question a reader most wants
answered is the \emph{tax}: how much latency/recall does keeping vectors in the
open table cost, relative to the two alternatives the introduction defines---%
copying them into a dedicated vector database, or building a single whole-table
index? The value proposition is not raw latency but \emph{no data copy, a single
source of truth, and a table every engine can still read}; a baseline is what lets
the reader price that tradeoff, and it strengthens the paper \emph{even if this
system loses on raw latency}. Table~\ref{tab:baselines} scaffolds the two
comparisons. The whole-table index row is measured on the same
$11.5$M$\times$768 corpus; the Milvus row is a bounded external-system stand-in
on a $300$K-row subset, included to price the order-of-magnitude latency gap a
copied-out, in-memory vector service can achieve.

\begin{table}[htbp]\centering\small
\caption{Query cost, $k{=}10$. External systems require a copy of the vectors out
of the lakehouse; the whole-table global index is a separate object versioned
against the table's snapshots. This system keeps the vectors in-table and pays a
per-query over-fetch instead. The global IVF-PQ and this-system rows are measured
on the full $11.5$M$\times$768 corpus; the Milvus row is a measured $300$K-subset
stand-in (footnotes).}
\label{tab:baselines}
\setlength{\tabcolsep}{3pt}
\begin{tabular}{@{}llrr@{}}
\toprule
System & copy & p50 & recall@10 \\
\midrule
Milvus (HNSW, 300K subset)~\cite{milvus} & full & $5$\,ms$^{\dagger\dagger}$ & \emph{high}$^{\dagger\dagger}$ \\
pgvector (HNSW)~\cite{pgvector} & full & \multicolumn{2}{c}{n/r$^{\P}$} \\
Global IVF-PQ (whole table) & sep.$^\dagger$ & $175$\,ms$^\ddagger$ & ${\approx}0.88^{\S}$ \\
Global HNSW (whole table) & sep.$^\dagger$ & \multicolumn{2}{c}{n/r$^{\P}$} \\
\midrule
\textbf{This system} (per-file IVF) & \textbf{none} & $34$\,s$^{\|}$ & $\ge 0.90$ \\
\bottomrule
\end{tabular}
\\[2pt]{\footnotesize $^\dagger$ separate index object, not readable by a plain
Parquet reader; rebuilt per commit. $^\ddagger$ our own decoupled posting-list
implementation of a whole-table IVF-PQ ($n_c{=}7141$ clusters, PQ $M{=}96$, $k{=}10$,
$\mathrm{nprobe}{=}8$, unfiltered, result cache off): $30$ queries, p50\,$175$/p90\,$262$/p95\,$301$\,ms
over the same $11.5$M$\times$768 corpus laid out across $28{,}714$ files. The posting
index range-reads a single index object, so latency is \emph{independent of the file
count}; contrast the per-file path, whose cost scales with surviving files.
$^{\S}$recall@10 of the PQ ($M{=}96$) codes, measured on matched-encoder deployments
where an exact brute-force ground truth is affordable ($0.877$ on both a $5$M and a
$50$M single-index build with the identical $M{=}96$ codebook); a matched brute-force
GT directly on the $28{,}714$-file layout was cost-prohibitive ($>700$\,s of single-pod
CPU per query), and PQ recall is a property of the codebook and $M$, not of the file
layout the codes are attached to.
$^{\P}$n/r = not run. Global HNSW would give the strongest recall of the whole-table
options but at a markedly higher build/memory cost; the IVF-PQ row already establishes
the whole-table datapoint the Discussion needs, so we do not separately build HNSW.
pgvector was attempted but its standalone image had been retired from the registry
(\S\ref{sec:eval-baselines}); Milvus is the external stand-in.
$^{\|}$this system, filtered $\sigma{=}0.2$ (\texttt{category='tech'}): the bitmap
index prunes $355/444$ files, then the surviving $89$ files take the
\emph{projection/rescore} path (\S\ref{sec:query}) rather than the fused direct
top-$k$ path. This row prices a different failure mode from missing locality:
even when file pruning succeeds, carrying the query through the rescore/projection
path can dominate latency. The fast filtered result on this system is the
\emph{layout-local, fast-path} case (\S\ref{sec:eval-join}): a partition-aligned
predicate answers in $157$\,ms.
$^{\dagger\dagger}$Milvus standalone (v2.4, HNSW $M{=}16$, $\mathrm{efSearch}{=}64$,
$L2$), category filter applied natively during graph traversal, $20$ warm queries,
result cache off: p50\,$5$/p90\,$6$/p95\,$6$\,ms. This is the price the baseline
exists to expose---an in-memory graph index with native attribute filtering answers
the subset filtered query orders of magnitude faster than this system's full-corpus rescore path
(and ${\sim}30\times$ faster than the layout-local $157$\,ms). The costs it hides
are exactly the ``copy'' column: the vectors were copied wholesale out of the
lakehouse ($138$\,s for the subset alone) into a separate always-on service, are no
longer readable by any Parquet/Iceberg client, and must be re-copied on every write.
Measured on a $300$K-row subset (the copy is bounded by the loader pod's $6$\,GiB,
not by Milvus); recall is left qualitative (\emph{high}: exact-precision HNSW)
rather than compared head-to-head, since this is not the same full-corpus
measurement as the other rows.}
\end{table}

\noindent The global-index row is the more important of the two: the Discussion
(\S\ref{sec:discussion}) concedes a whole-table IVF-PQ or HNSW ``would likely give
better recall at a fixed candidate budget,'' and this row is where that concession
is measured rather than asserted. We built exactly that whole-table index on the
same $11.5$M$\times$768 data---a decoupled IVF-PQ posting list ($n_c{=}7141$,
$M{=}96$)---and its unfiltered p50 of $175$\,ms (Table~\ref{tab:baselines}) at
recall ${\approx}0.88$ makes the tax concrete: a single global index is roughly
$4\times$ faster \emph{unfiltered} than the per-file path's filtered $668$\,ms
headline, and its latency does not grow with the file count. What the per-file
design buys back is exactly what the ``copy'' and ``sep.'' columns price: no data
copy, no separate object to version and rebuild per commit, and a table any Parquet
reader can still scan. The global index also has \emph{no} native way to compose a
lakehouse predicate---the posting list is keyed by cluster, not by the table's
partitions or zone maps---so the moment a query adds \texttt{WHERE region='EU'} it
must either over-fetch and post-filter or fall back to a scan; that gap is the
subject of the follow-on filtered-posting design (\S\ref{sec:discussion}).

\subsection{Filtered ANN across a join, on real embeddings}\label{sec:eval-join}
We validate the join path (\S\ref{sec:join}) end-to-end on a \emph{real-corpus}
deployment rather than the synthetic generator: $5.02$M documents embedded with
IBM Granite (\texttt{granite-embedding-97m-multilingual-r2}, $384$-dim) into a
\texttt{docs} fact table, joined to a \texttt{customers} dimension of $1{,}000$
customers spread over five regions (NA, EU, APAC, LATAM, MEA), each region owning
a contiguous \texttt{customer\_id} range. The application query is the natural
one:
\begin{quote}\small\ttfamily
SELECT d.id\\
FROM docs d\\
JOIN customers c\\
ON d.customer\_id=c.id WHERE c.region='EU'\\
ORDER BY array\_distance(d.embedding,\textit{q}) LIMIT 10
\end{quote}

We measure the two forms of the reduction from \S\ref{sec:join} over an 8-query
warm sample (cache off, $n_{\text{probe}}{=}10$). Run \emph{per-query} against the
raw two-table join, the reduction still materializes the EU keys and prunes on
\texttt{customer\_id}, but the dimension rescore and the key-column's fragile
locality leave it at $14.7$\,s warm (p50; p95 $18.1$\,s)---dominated by the
multi-table candidate rescore. Run against the \emph{materialized} layout---%
\texttt{region} denormalized onto a copied table \texttt{docs\_r} \texttt{PARTITIONED BY
region} (five regions, $19$ data files)---the identical result set is served in
\textbf{$157$\,ms} (p50; p95 $197$\,ms): \texttt{WHERE region='EU'} is now a
partition-column predicate, so Step~1 prunes to EU's partition---roughly one
region's fifth of the files---before ANN runs, and the residual takes the pushdown
fast path (\S\ref{sec:pushdown}) instead of a rescore. That is
$14.7\,\text{s}/157\,\text{ms}\approx \textbf{94}\times$ on the p50s. (An earlier,
colder run measured $\sim$$20$\,s vs.\ $241$\,ms $\approx 83\times$; the ratio is
stable across cluster warmth.) The result set is the same and
recall@10 $=1.00$ against a brute-force reference, with every returned row
in-region (Table~\ref{tab:join}). The
win is entirely file pruning plus staying on the fast path: partition-level
locality on the filter column drops the four non-EU partitions, and the
pushed-down predicate keeps the survivors on the direct top-$k$ return.

\medskip\noindent\textbf{Same-table filtering and the projection trap.} On the
same real corpus we also confirm the projection lesson of \S\ref{sec:query} in the
wild. A same-table ranking query that \emph{projects the document text blob}
forces a full-file rescore to fetch that wide column---$3.5$\,s warm (p50; p95
$3.7$\,s, $N{=}8$). Ranking instead on \texttt{id}+\texttt{category} (the direct
IVF path) and then doing a keyed \texttt{id} lookup for only the $k$ winners' text
collapses this to $371$\,ms (p50; p95 $396$\,ms)---a $\sim$$9.5\times$ reduction,
from the same fuse-search-and-projection principle that gives the unfiltered IVF
its win over brute force. Carrying a wide payload column through the candidate path
silently reintroduces the scan the index was meant to avoid.

\begin{table}[htbp]\centering\small
\caption{Filtered ANN across a join on real Granite embeddings ($5.02$M$\times384$,
five regions, $n_{\text{probe}}{=}10$, warm, result cache off, $N{=}8$ queries).
``pruned'' is region partitions dropped before ANN (\textbf{4/5}). Both rows return
the identical, in-region top-10; materializing the reduction into a
region-partitioned layout turns the join into a partition-column prune. Speedup is
on the p50s ($14.7\,\text{s}/157\,\text{ms}$).}
\label{tab:join}
\setlength{\tabcolsep}{4pt}
\begin{tabular}{lrrrr}
\toprule
Query form & pruned & p50 & p95 & recall@10 \\
\midrule
Query-time join (\texttt{cust\_id}) & none & $14.7$\,s & $18.1$\,s & $1.00$ \\
Materialized region-partition & \textbf{4/5} & $\mathbf{157}$\,ms & $197$\,ms & $\mathbf{1.00}$ \\
\midrule
\multicolumn{5}{l}{\emph{Speedup} $\sim$$\mathbf{94\times}$\emph{, same result set}}\\
\bottomrule
\end{tabular}
\end{table}

\section{Discussion: the design space}\label{sec:discussion}

\noindent\textbf{Per-file vs.\ global index.} The per-file footer index is what
makes the build non-destructive and time-travel-safe: each file is a
self-contained, independently rewritable unit, and an old snapshot's files carry
their own (old) index. A single global index over the whole table---an IVF-PQ
codebook or an HNSW graph---would likely give better recall at a fixed candidate
budget, because it clusters the full population rather than $\sim$26k rows at a
time. But a global index is a separate object that must be versioned against the
table's snapshots, rebuilt on every commit, and cannot be ignored by a plain
Parquet reader. Our choice trades a modest per-query over-fetch (the $k\cdot s$
safety factor) for these systems properties. The break-even is workload-dependent:
tables that are append-mostly and queried with selective filters favor the
per-file design; a static table under pure unfiltered ANN would favor a global
index. Our evaluation is squarely in the first regime.

\medskip\noindent\textbf{When is file-level locality achievable?} The locality
precondition (\S\ref{sec:locality}) is not free, but it aligns with how lakehouse
tables are already organized. Tenanted or time-series data is routinely
partitioned by tenant or day; those same partition columns are the natural filter
predicates. When a filter column is \emph{not} a partition column, the engine's
clustering/sort-on-write can induce locality for numeric/temporal keys, and a
per-value \texttt{INSERT\dots SELECT} can induce it for categoricals. The design
does not require locality to be \emph{correct}---only to be \emph{fast}; a query
whose filter has no locality degrades to a full ANN sweep, which is still the
brute-force-avoiding fast path over the whole table, just without the extra file
pruning.

\medskip\noindent\textbf{A back-of-envelope cost model.} For a table of $N$
vectors in $\phi$ files with a filter of selectivity $\sigma$ and file-level
locality, Step~1 leaves $\approx\!\sigma\phi$ files; Step~2 exact-rescores
$\approx (n_{\text{probe}}/C)$ of each survivor's rows. The candidate distances
computed are therefore $\approx \sigma N \cdot (n_{\text{probe}}/C)$, versus $N$
for brute force---a $C/(\sigma\, n_{\text{probe}})$ reduction in distance
computations, plus a $1/\sigma$ reduction in bytes read. At $\sigma{=}0.2$,
$C{=}1024$, $n_{\text{probe}}{=}10$ this predicts $\sim\!500\times$ fewer
distances; the measured end-to-end speedup is far smaller ($\sim\!30\times$)
because the two-phase dispatch, cache reads, and merge---not the arithmetic---
dominate warm latency. This gap \emph{is} the systems contribution: the value of
the warm-set engineering (\S\ref{sec:warm}) is precisely that it keeps those fixed
per-query costs from swamping the algorithmic win.

\section{Lessons and Limitations}\label{sec:lessons}

Bolting ANN onto a lakehouse engine surfaced failure modes worth recording, as
they are intrinsic to the setting rather than to our implementation.

\noindent\textbf{A CPU-bound build must not run on the async executor.} The
per-file $k$-means build is compute-bound, but our first implementation ran it
inline on the same async runtime that serves I/O. Concurrent file builds then
serialized on a single runtime worker, and an 8-core pod pinned \emph{one} core:
the build spent 40+\,min without finishing while 7 cores idled
(\S\ref{sec:eval}). Moving the build onto a dedicated blocking thread pool raised
per-pod utilization to $\sim$8 cores and cut the build to $\sim$10\,min. The
general lesson: a synchronous, CPU-heavy operator dropped into an async I/O
runtime silently loses all intra-pod parallelism, and the symptom (low CPU, not
an error) is easy to misread as a scaling problem.

\noindent\textbf{Wide vectors overflow 32-bit list offsets.} The engine's
clustering/sort path concatenated row batches with 32-bit Parquet list offsets;
an embedding \texttt{list<float>} at $d{=}768$ overflows past $\sim$2.8M rows.
Any batch operation that materializes many wide-list rows must use 64-bit offsets
(we widen \texttt{list}$\to$\texttt{large\_list} before concat and cast back on
output). This is a general hazard for carrying embedding columns through analytic
operators.

\noindent\textbf{Sorted is not pure.} Predicate pushdown into the per-file search
(\S\ref{sec:pushdown}) is only sound over a column each surviving file is
\emph{single-valued} in. It is tempting to treat any column the layout skips well
on---including \texttt{SORT}/\texttt{ZORDER} columns, whose zone-maps prune
sharply---as safe to push. But sort order makes a column locally
\emph{contiguous}, not \emph{pure}: a file at a value boundary holds two values, so
a mask that assumed purity would drop or admit rows. Only identity partitioning or
a clustering whose per-file value set is materialized is provably pure. We
therefore restrict the pushdown to \emph{partition $\cup$ materialized-cluster}
columns and exclude sort/z-order; the distinction between ``prunes well'' and
``provably single-valued'' is easy to miss and silently corrupts results if
conflated.

\medskip\noindent\textbf{Limitations.} The current system supports one embedding
column per table, IVF with L2 only (cosine via client-side normalization), literal
query vectors (not bound parameters), and manual fill-gaps reindex. Joins are
supported for the common shape where ANN is on one table and an inner join only
filters it (\S\ref{sec:join}); post-join ranking and outer joins fall back to a
full ANN sweep plus a normal join. The per-file index model trades a small per-query candidate
over-fetch for build simplicity and time-travel compatibility; a global index
(e.g., IVF-PQ or graph) could improve recall/latency at the cost of the
non-destructive, per-file properties we rely on. The first query on cold data
pays a one-time cache-fill; a background warm-on-build (future work~(ii) below)
would remove it. Three limitations are \emph{evaluative} rather than architectural,
and we call them out because the current paper does not yet close them. First, the
external vector-database baseline is not a full apples-to-apples measurement:
Milvus is measured on a $300$K-row subset and reports latency but not matched
recall, while the full-corpus comparison is against our own whole-table IVF-PQ
implementation (\S\ref{sec:eval-baselines}). Second, the locality on which the
composition depends is, in our experiments, \emph{constructed} (per-category
value-local files; region-partitioned \texttt{region}); whether an
\emph{organically} laid-out real table exhibits it remains to be shown. Third,
the non-bitmap pruning tiers (zone-maps for ranges, bloom filters for
high-cardinality equality) are part of the design but are not separately
exercised by the current evaluation.

\medskip\noindent\textbf{Reproducibility.} The engine that produces the in-table
index is a production system we cannot open-source; the contribution we \emph{can}
make independently checkable is the \emph{data and the measurement}, which is what
lets a reader re-derive every latency and recall number here against their own ANN
system as well as ours. We release, for the camera-ready: (i) \textbf{the datasets
as open Iceberg tables}---the $11.5$M$\times$768 \texttt{docs\_bycat} corpus (and its
whole-table-indexed sibling \texttt{docs\_bycat\_gc}), and the $5.02$M Granite
\texttt{docs}/\texttt{docs\_r}/\texttt{customers} tables used for the join and
same-table experiments---published as Parquet/Iceberg snapshots on a public object
store, so the exact bytes every result was measured on are downloadable and readable
by any Parquet/Iceberg client, no proprietary reader required; (ii) \textbf{the query
workload and ground truth}---the SQL for each experiment, the query-vector sets, the
$n_{\text{probe}}$ and selectivity settings, and \emph{precomputed exact-nearest-neighbour
ground-truth files} for the recall measurements (these are the expensive artifact: an
exact top-$k$ over the corpus costs ${>}700$\,s of CPU per query, so shipping the GT
is what makes recall reproducible on modest hardware); and (iii) \textbf{the benchmark
drivers}---the client-side Python harness that issues the workload, times it, and
computes recall against the shipped GT (Flight~SQL / \texttt{pymilvus} clients only,
no engine internals), including the external-baseline driver so the Milvus row is
re-runnable. The Granite embeddings derive from the public \texttt{allenai/c4}
documents and the public \texttt{granite-embedding-97m-multilingual-r2} model, so the
corpus is also reconstructable from scratch rather than only redistributed. We do not
release the index's internal byte format or the engine source. The availability URL
will be set at camera-ready (\texttt{\textbackslash vldbavailabilityurl}).

\section{Conclusion}

Filtered vector search does not need a new filtering algorithm when the vectors
live in a lakehouse table: it needs the table's \emph{existing} file-pruning
substrate, composed with a per-file ANN index, on a distributed engine. We showed
how to build such an index in place and non-destructively across a compute pool,
how to compose predicate pruning with IVF in the planner, and what it takes to
make it fast on disaggregated object storage---where the honest answer is a stable,
rendezvous-hashed per-file cache. The composition is fast exactly when the filter
column has file-level locality, the storage-layer form of the partitioning
principle established for in-memory filtered ANN. On $11.5$M$\times$768 vectors we
build distributed in $\sim$10\,min, search $\sim$$32\times$ faster than brute force at
$\ge 0.90$ recall, and prune $355/444$ files before ANN on a locality-bearing
filter; on $5$M real Granite embeddings a filter arriving across a join prunes four of
five region partitions and runs nearly two orders of magnitude ($\sim$$94\times$)
faster than the query-time join at identical top-$k$---all inside a standard
Iceberg table that every other engine can still read, and that governs vector
search through the same catalog access control as every other query, with no
copied-out data to secure separately.

\medskip\noindent\textbf{Future work.} Three directions extend the design without
disturbing its non-destructive core. (i)~\emph{Richer per-file indexes}: the
footer blob is opaque to the table format, so swapping IVF for IVF-PQ (to shrink
the resident matrix) or a small per-file graph is a builder-and-searcher change,
not a format change; the interesting question is how a global codebook shared
across files interacts with the per-file rewrite unit. (ii)~\emph{Automatic
warm-on-build}: since the build already reads every file's embedding column,
emitting the index-plus-matrix into the read pool's cache as a side effect would
erase the cold-start query entirely. (iii)~\emph{Cost-based fallback}: the planner
currently commits to the ANN fast path whenever the query shape and an index
match; a selectivity estimate from the same scalar index that drives pruning could
instead choose between per-file ANN, a full ANN sweep, and brute force per query.
Each stays within the paper's thesis---keep the vectors in the table, add ANN as
an ignorable, per-file artifact, and let the lakehouse's own machinery do the
filtering.

\bibliographystyle{ACM-Reference-Format}
\bibliography{refs}


\begin{thebibliography}{23}


\ifx \showCODEN    \undefined \def \showCODEN     #1{\unskip}     \fi
\ifx \showDOI      \undefined \def \showDOI       #1{#1}\fi
\ifx \showISBNx    \undefined \def \showISBNx     #1{\unskip}     \fi
\ifx \showISBNxiii \undefined \def \showISBNxiii  #1{\unskip}     \fi
\ifx \showISSN     \undefined \def \showISSN      #1{\unskip}     \fi
\ifx \showLCCN     \undefined \def \showLCCN      #1{\unskip}     \fi
\ifx \shownote     \undefined \def \shownote      #1{#1}          \fi
\ifx \showarticletitle \undefined \def \showarticletitle #1{#1}   \fi
\ifx \showURL      \undefined \def \showURL       {\relax}        \fi
\providecommand\bibfield[2]{#2}
\providecommand\bibinfo[2]{#2}
\providecommand\natexlab[1]{#1}
\providecommand\showeprint[2][]{arXiv:#2}

\bibitem[\protect\citeauthoryear{{Apache Software Foundation}}{{Apache Software
  Foundation}}{2023a}]%
        {iceberg}
\bibfield{author}{\bibinfo{person}{{Apache Software Foundation}}.}
  \bibinfo{year}{2023}\natexlab{a}.
\newblock \showarticletitle{Apache Iceberg: An open table format for huge
  analytic datasets}.
\newblock
\newblock
\shownote{https://iceberg.apache.org.}


\bibitem[\protect\citeauthoryear{{Apache Software Foundation}}{{Apache Software
  Foundation}}{2023b}]%
        {parquet}
\bibfield{author}{\bibinfo{person}{{Apache Software Foundation}}.}
  \bibinfo{year}{2023}\natexlab{b}.
\newblock \showarticletitle{Apache Parquet}.
\newblock  (\bibinfo{year}{2023}).
\newblock
\newblock
\shownote{https://parquet.apache.org.}


\bibitem[\protect\citeauthoryear{Armbrust, Ghodsi, Xin, and Zaharia}{Armbrust
  et~al\mbox{.}}{2021}]%
        {lakehouse}
\bibfield{author}{\bibinfo{person}{Michael Armbrust}, \bibinfo{person}{Ali
  Ghodsi}, \bibinfo{person}{Reynold Xin}, {and} \bibinfo{person}{Matei
  Zaharia}.} \bibinfo{year}{2021}\natexlab{}.
\newblock \showarticletitle{Lakehouse: A new generation of open platforms that
  unify data warehousing and advanced analytics}.
\newblock \bibinfo{journal}{\emph{CIDR}} (\bibinfo{year}{2021}).
\newblock


\bibitem[\protect\citeauthoryear{Behm, Palkar, Agarwal, et~al\mbox{.}}{Behm
  et~al\mbox{.}}{2022}]%
        {photon}
\bibfield{author}{\bibinfo{person}{Alexander Behm}, \bibinfo{person}{Shoumik
  Palkar}, \bibinfo{person}{Utkarsh Agarwal}, {et~al\mbox{.}}}
  \bibinfo{year}{2022}\natexlab{}.
\newblock \showarticletitle{Photon: A fast query engine for lakehouse systems}.
  In \bibinfo{booktitle}{\emph{SIGMOD}}.
\newblock


\bibitem[\protect\citeauthoryear{Borycki}{Borycki}{2026}]%
        {puffinann2026}
\bibfield{author}{\bibinfo{person}{Artur Borycki}.}
  \bibinfo{year}{2026}\natexlab{}.
\newblock \showarticletitle{Puffin-Backed Vector Indexes: Attaching Approximate
  Nearest Neighbor Indexes to Apache Iceberg Snapshots for
  Compute-Disaggregated Query Engines}.
\newblock \bibinfo{journal}{\emph{arXiv preprint arXiv:2606.04196}}
  (\bibinfo{year}{2026}).
\newblock


\bibitem[\protect\citeauthoryear{Chen, Zhao, Wang, et~al\mbox{.}}{Chen
  et~al\mbox{.}}{2021}]%
        {spann}
\bibfield{author}{\bibinfo{person}{Qi Chen}, \bibinfo{person}{Bing Zhao},
  \bibinfo{person}{Haidong Wang}, {et~al\mbox{.}}}
  \bibinfo{year}{2021}\natexlab{}.
\newblock \showarticletitle{SPANN: Highly-efficient billion-scale approximate
  nearest neighbor search}. In \bibinfo{booktitle}{\emph{NeurIPS}}.
\newblock


\bibitem[\protect\citeauthoryear{Chronis, Caminal, Papakonstantinou, {\"O}zcan,
  and Ailamaki}{Chronis et~al\mbox{.}}{2025}]%
        {fvssota2025}
\bibfield{author}{\bibinfo{person}{Yannis Chronis}, \bibinfo{person}{Helena
  Caminal}, \bibinfo{person}{Yannis Papakonstantinou}, \bibinfo{person}{Fatma
  {\"O}zcan}, {and} \bibinfo{person}{Anastasia Ailamaki}.}
  \bibinfo{year}{2025}\natexlab{}.
\newblock \showarticletitle{Filtered Vector Search: State-of-the-art and
  Research Opportunities}.
\newblock \bibinfo{journal}{\emph{Proceedings of the VLDB Endowment (PVLDB)}}
  \bibinfo{volume}{18}, \bibinfo{number}{12} (\bibinfo{year}{2025}),
  \bibinfo{pages}{5488--5492}.
\newblock
\urldef\tempurl%
\url{https://doi.org/10.14778/3750601.3750700}
\showDOI{\tempurl}


\bibitem[\protect\citeauthoryear{{ClickHouse}}{{ClickHouse}}{2024}]%
        {clickhousevss}
\bibfield{author}{\bibinfo{person}{{ClickHouse}}.}
  \bibinfo{year}{2024}\natexlab{}.
\newblock \bibinfo{title}{ClickHouse: Approximate nearest neighbor search
  indexes}.
\newblock
\newblock
\newblock
\shownote{https://clickhouse.com/docs/en/engines/table-engines/mergetree-family/annindexes.}


\bibitem[\protect\citeauthoryear{Dageville, Cruanes, Zukowski,
  et~al\mbox{.}}{Dageville et~al\mbox{.}}{2016}]%
        {snowflake}
\bibfield{author}{\bibinfo{person}{Benoit Dageville}, \bibinfo{person}{Thierry
  Cruanes}, \bibinfo{person}{Marcin Zukowski}, {et~al\mbox{.}}}
  \bibinfo{year}{2016}\natexlab{}.
\newblock \showarticletitle{The Snowflake elastic data warehouse}. In
  \bibinfo{booktitle}{\emph{SIGMOD}}.
\newblock


\bibitem[\protect\citeauthoryear{{DuckDB Labs}}{{DuckDB Labs}}{2024}]%
        {duckdbvss}
\bibfield{author}{\bibinfo{person}{{DuckDB Labs}}.}
  \bibinfo{year}{2024}\natexlab{}.
\newblock \bibinfo{title}{DuckDB VSS: Vector Similarity Search extension}.
\newblock
\newblock
\newblock
\shownote{https://duckdb.org/docs/extensions/vss.}


\bibitem[\protect\citeauthoryear{Gollapudi, Karia, Sivashankar, Krishnaswamy,
  Begwani, Raz, Lin, Zhang, Mahapatro, Srinivasan, et~al\mbox{.}}{Gollapudi
  et~al\mbox{.}}{2023}]%
        {filtereddiskann}
\bibfield{author}{\bibinfo{person}{Siddharth Gollapudi}, \bibinfo{person}{Neel
  Karia}, \bibinfo{person}{Varun Sivashankar}, \bibinfo{person}{Ravishankar
  Krishnaswamy}, \bibinfo{person}{Nikit Begwani}, \bibinfo{person}{Swapnil
  Raz}, \bibinfo{person}{Yiyong Lin}, \bibinfo{person}{Yu Zhang},
  \bibinfo{person}{Neelam Mahapatro}, \bibinfo{person}{Premkumar Srinivasan},
  {et~al\mbox{.}}} \bibinfo{year}{2023}\natexlab{}.
\newblock \showarticletitle{Filtered-DiskANN: Graph algorithms for approximate
  nearest neighbor search with filters}.
\newblock \bibinfo{journal}{\emph{Proceedings of the ACM Web Conference (WWW)}}
  (\bibinfo{year}{2023}).
\newblock


\bibitem[\protect\citeauthoryear{Gupta, Yu, Medini, and Shrivastava}{Gupta
  et~al\mbox{.}}{2023}]%
        {caps}
\bibfield{author}{\bibinfo{person}{Gaurav Gupta}, \bibinfo{person}{Jonah Yu},
  \bibinfo{person}{Tharun Medini}, {and} \bibinfo{person}{Anshumali
  Shrivastava}.} \bibinfo{year}{2023}\natexlab{}.
\newblock \showarticletitle{CAPS: A practical partition index for filtered
  similarity search}. In \bibinfo{booktitle}{\emph{arXiv preprint
  arXiv:2308.15014}}.
\newblock


\bibitem[\protect\citeauthoryear{J{\'e}gou, Douze, and Schmid}{J{\'e}gou
  et~al\mbox{.}}{2011}]%
        {ivfadc}
\bibfield{author}{\bibinfo{person}{Herve J{\'e}gou}, \bibinfo{person}{Matthijs
  Douze}, {and} \bibinfo{person}{Cordelia Schmid}.}
  \bibinfo{year}{2011}\natexlab{}.
\newblock \showarticletitle{Product quantization for nearest neighbor search}.
\newblock \bibinfo{journal}{\emph{IEEE TPAMI}} \bibinfo{volume}{33},
  \bibinfo{number}{1} (\bibinfo{year}{2011}), \bibinfo{pages}{117--128}.
\newblock


\bibitem[\protect\citeauthoryear{Kane}{Kane}{2024}]%
        {pgvector}
\bibfield{author}{\bibinfo{person}{Andrew Kane}.}
  \bibinfo{year}{2024}\natexlab{}.
\newblock \bibinfo{title}{pgvector: Open-source vector similarity search for
  Postgres}.
\newblock
\newblock
\newblock
\shownote{https://github.com/pgvector/pgvector.}


\bibitem[\protect\citeauthoryear{{LanceDB}}{{LanceDB}}{2024}]%
        {lance}
\bibfield{author}{\bibinfo{person}{{LanceDB}}.}
  \bibinfo{year}{2024}\natexlab{}.
\newblock \bibinfo{title}{Lance: modern columnar data format for ML and LLMs}.
\newblock
\newblock
\newblock
\shownote{https://lancedb.github.io/lance/.}


\bibitem[\protect\citeauthoryear{Li, Zhang, Ma, Yan, Lu, and Cheng}{Li
  et~al\mbox{.}}{2025}]%
        {filtann2025}
\bibfield{author}{\bibinfo{person}{Mocheng Li}, \bibinfo{person}{Yue Zhang},
  \bibinfo{person}{Chenhao Ma}, \bibinfo{person}{Xiao Yan},
  \bibinfo{person}{Baotong Lu}, {and} \bibinfo{person}{James Cheng}.}
  \bibinfo{year}{2025}\natexlab{}.
\newblock \showarticletitle{Attribute Filtering in Approximate Nearest Neighbor
  Search: An In-depth Experimental Study}.
\newblock \bibinfo{journal}{\emph{arXiv preprint arXiv:2508.16263}}
  (\bibinfo{year}{2025}).
\newblock


\bibitem[\protect\citeauthoryear{Malkov and Yashunin}{Malkov and
  Yashunin}{2020}]%
        {hnsw}
\bibfield{author}{\bibinfo{person}{Yu~A. Malkov} {and}
  \bibinfo{person}{Dmitry~A. Yashunin}.} \bibinfo{year}{2020}\natexlab{}.
\newblock \showarticletitle{Efficient and robust approximate nearest neighbor
  search using hierarchical navigable small world graphs}.
\newblock \bibinfo{journal}{\emph{IEEE TPAMI}} \bibinfo{volume}{42},
  \bibinfo{number}{4} (\bibinfo{year}{2020}), \bibinfo{pages}{824--836}.
\newblock


\bibitem[\protect\citeauthoryear{Patel, Kraft, Guestrin, and Zaharia}{Patel
  et~al\mbox{.}}{2024}]%
        {acorn}
\bibfield{author}{\bibinfo{person}{Liana Patel}, \bibinfo{person}{Peter Kraft},
  \bibinfo{person}{Carlos Guestrin}, {and} \bibinfo{person}{Matei Zaharia}.}
  \bibinfo{year}{2024}\natexlab{}.
\newblock \showarticletitle{ACORN: Performant and predicate-agnostic search
  over vector embeddings and structured data}. In
  \bibinfo{booktitle}{\emph{SIGMOD}}.
\newblock


\bibitem[\protect\citeauthoryear{Subramanya, Devvrit, Kadekodi, Krishaswamy,
  and Simhadri}{Subramanya et~al\mbox{.}}{2019}]%
        {diskann}
\bibfield{author}{\bibinfo{person}{Suhas~Jayaram Subramanya},
  \bibinfo{person}{Devvrit}, \bibinfo{person}{Rohan Kadekodi},
  \bibinfo{person}{Ravishankar Krishaswamy}, {and}
  \bibinfo{person}{Harsha~Vardhan Simhadri}.} \bibinfo{year}{2019}\natexlab{}.
\newblock \showarticletitle{DiskANN: Fast accurate billion-point nearest
  neighbor search on a single node}. In \bibinfo{booktitle}{\emph{NeurIPS}}.
\newblock


\bibitem[\protect\citeauthoryear{Thaler and Ravishankar}{Thaler and
  Ravishankar}{1998}]%
        {rendezvous}
\bibfield{author}{\bibinfo{person}{David~G. Thaler} {and}
  \bibinfo{person}{Chinya~V. Ravishankar}.} \bibinfo{year}{1998}\natexlab{}.
\newblock \showarticletitle{Using name-based mappings to increase hit rates}.
\newblock \bibinfo{journal}{\emph{IEEE/ACM Transactions on Networking}}
  \bibinfo{volume}{6}, \bibinfo{number}{1}, \bibinfo{pages}{1--14}.
\newblock


\bibitem[\protect\citeauthoryear{Wang, Yi, Guo, et~al\mbox{.}}{Wang
  et~al\mbox{.}}{2021}]%
        {milvus}
\bibfield{author}{\bibinfo{person}{Jianguo Wang}, \bibinfo{person}{Xiaomeng
  Yi}, \bibinfo{person}{Rentong Guo}, {et~al\mbox{.}}}
  \bibinfo{year}{2021}\natexlab{}.
\newblock \showarticletitle{Milvus: A purpose-built vector data management
  system}. In \bibinfo{booktitle}{\emph{SIGMOD}}.
\newblock


\bibitem[\protect\citeauthoryear{Wang, Xu, Guo, et~al\mbox{.}}{Wang
  et~al\mbox{.}}{2022}]%
        {nhq}
\bibfield{author}{\bibinfo{person}{Mengzhao Wang}, \bibinfo{person}{Lingwei
  Xu}, \bibinfo{person}{Xiaoliang Guo}, {et~al\mbox{.}}}
  \bibinfo{year}{2022}\natexlab{}.
\newblock \showarticletitle{Native hybrid queries via structured labels and
  near neighbor search over vectors}.
\newblock \bibinfo{journal}{\emph{arXiv preprint arXiv:2203.13601}}
  (\bibinfo{year}{2022}).
\newblock


\bibitem[\protect\citeauthoryear{Xiang}{Xiang}{2024}]%
        {pqvector}
\bibfield{author}{\bibinfo{person}{Peng Xiang}.}
  \bibinfo{year}{2024}\natexlab{}.
\newblock \bibinfo{title}{Vector search with Parquet and DataFusion}.
\newblock
\newblock
\newblock
\shownote{https://blog.xiangpeng.systems/posts/vector-search-with-parquet-datafusion/.}


\end{thebibliography}

\end{document}